\documentclass[conference,a4paper]{IEEEtran}
\usepackage{graphicx,xcolor}
\usepackage{subfig}
\usepackage{float}
\usepackage{cite}
\usepackage{amsmath}
\ifCLASSINFOpdf
\else
\fi

\usepackage[left=1.41cm,right=1.44cm,top=1.91cm,bottom=4.2cm,bindingoffset=0cm]{geometry}
\usepackage{multirow}
\DeclareRobustCommand*{\IEEEauthorrefmark}[1]{\raisebox{0pt}[0pt][0pt]{\textsuperscript{\footnotesize #1}}}

\begin{document}
\title{Path Loss Characterization for Intra-Vehicle Wearable Deployments at 60 GHz}

\author{\IEEEauthorblockN{
Vasilii Semkin\IEEEauthorrefmark{1,2},   
Aleksei Ponomarenko-Timofeev\IEEEauthorrefmark{2},   
Aki Karttunen\IEEEauthorrefmark{3},
Olga Galinina\IEEEauthorrefmark{2}, \\
Sergey Andreev\IEEEauthorrefmark{2}, and    
Yevgeni Koucheryavy\IEEEauthorrefmark{2}  
}                                     
\IEEEauthorblockA{\IEEEauthorrefmark{1}
Universite catholique de Louvain, Louvain-la-Neuve, Belgium \\vasilii.semkin@uclouvain.be}
\IEEEauthorblockA{\IEEEauthorrefmark{2}
Laboratory of Electronics and Communications Engineering, Tampere University of Technology, Finland \\\{aleksei.ponomarenko-timofeev, olga.galinina, sergey.andreev, evgeni.kucheryavy\}@tut.fi}
\IEEEauthorblockA{\IEEEauthorrefmark{3}
Department of Electronics and Nanoengineering, Aalto University, Espoo, Finland \\aki.karttunen@aalto.fi}
}

\maketitle

\begin{abstract}
In this work, we present the results of a wideband measurement campaign at $60$\,GHz conducted inside a Linkker electric city bus. Targeting prospective millimeter-wave (mmWave) public transportation wearable scenarios, we mimic a typical deployment of mobile high-end consumer devices in a dense environment. Specifically, our intra-vehicle deployment includes one receiver and multiple transmitters corresponding to a mmWave access point and passengers' wearable and hand-held devices. While the receiver is located in the front part of the bus, the transmitters repeat realistic locations of personal devices (i) at the seat level (e.g., a hand-held device) and (ii) at a height $70$\,cm above the seat (e.g., a wearable device: augmented reality glasses or a head-mounted display). Based on the measured received power, we construct a logarithmic model for the distance-dependent path loss. The parametrized models developed in the course of this study have the potential to become an attractive ground for the link budget estimation and interference footprint studies in crowded public transportation scenarios.
\end{abstract}


\IEEEpeerreviewmaketitle
\vspace{7pt}


\section{Introduction}

Contemporary wireless systems are driven by the evolution of novel bandwidth-hungry mobile applications, such as virtual reality, mobile gaming, and vehicular entertainment services, which require high transmission quality and low latency. One of the biggest challenges in terms of pervasive connectivity comes with the growing popularity and affordability of standalone high-end consumer wearable devices.

To meet the ever-growing data rate demands in densely populated scenarios such as, e.g., public transportation (suburban trains, metro, and buses), prospective service providers will need to rely on communication systems operating at millemeter-wave (mmWave) frequencies~\cite{venugopal_heath_wearables16}. One of the most promising choices for short-range mmWave communications~\cite{ghasempour2017ieee, Zhou2018} is the unlicensed $60$\,GHz-band controlled by, e.g., recent IEEE 802.11ay protocol~\cite{tg80211ay} that supports data rates of multiple Gbps. 

As a result, the path loss characterization at $60$\,GHz becomes a crucial component in wireless network planning and optimization. While the radio wave propagation at mmWave frequencies has been extensively studied in outdoor and office environments, the public transportation scenarios have not yet received much attention. Similar studies, however, covered channel propagation and modeling efforts for the lower frequency range. For example, in~\cite{moraitis_valtr_aircraft}, the authors presented a study of the radio wave propagation based on the path loss simulation and measurements in a civil aircraft at microwave frequencies. Characterization of $2.4$\,GHz channel inside buses was performed in~\cite{azpilicueta15_bus}. 

In this paper, we provide the results of our wideband measurements at $60$\,GHz frequency conducted inside an electric city bus. In~\cite{chandra16_PL_60G_bus}, the authors presented a similar study of the path loss characteristics at $60$\,GHz inside a touristic bus; however, the geometry of a typical city bus differs from that of the long-distance buses as in~\cite{chandra16_PL_60G_bus}. Our main contribution lies in constructing easy-to-implement and analytically tractable path loss models based on the real mmWave propagation data. We believe that our results have significant potential for the system-level evaluation, e.g., link budget estimation and interference footprint studies, as well as for network deployment planning in future public transport scenarios.

The paper is organized as follows. Section II briefly describes the employed wideband equipment and the scenario of interest. In Section III, we process the measurement results by designing a statistical model based on the logarithmic dependency between the mean path loss and the distance between the transmitter and the receiver. Finally, Section IV provides a short discussion of the results and concludes the paper.

\vspace{7pt}
\section{Measurement scenario}

\begin{figure}[!t]
  \centering
  \includegraphics[scale=0.8]{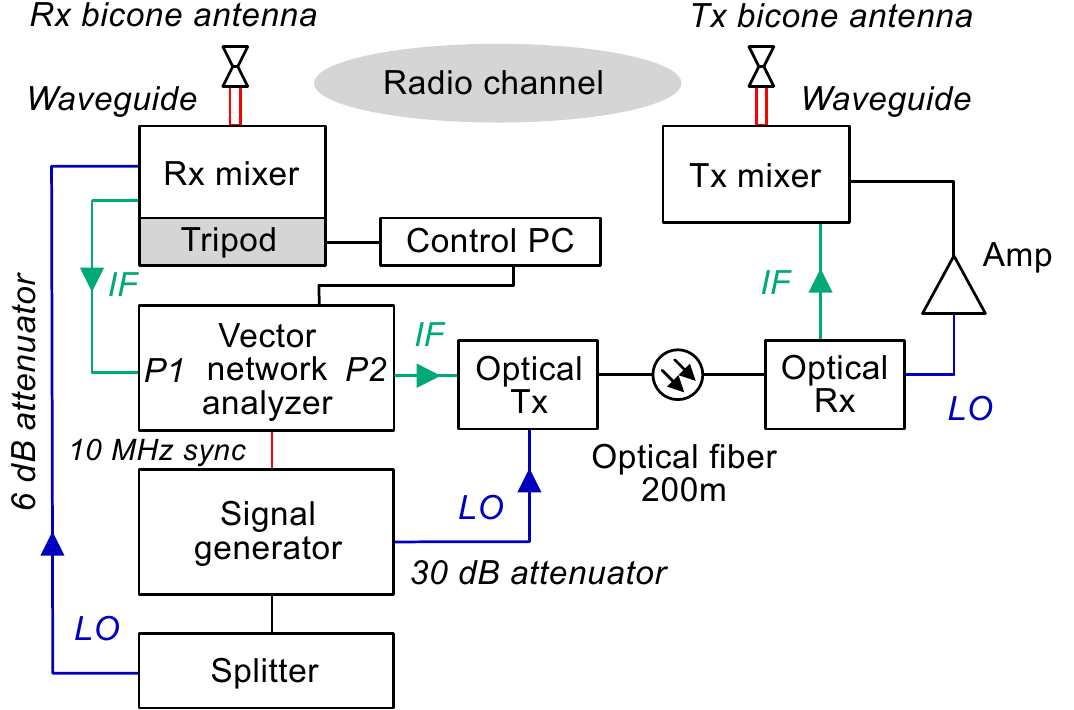}
  \caption{Block diagram of the measurement equipment. 
  }
  \label{fig:equip}
\end{figure}

Our experimental campaign has been carried out inside a Linkker electric bus\footnote{The first Finnish fast-charging electric bus, http://www.linkkerbus.com} of length $12.80$\,m and width $2.55$\,m. All measurements have been performed at $60$\,GHz frequency with the use of a wideband channel sounder. The employed equipment presented in Fig.~\ref{fig:equip} includes a vector network analyzer, a signal generator, up- and down- converters, and transmitting/receiving omnidirectional antennas of $2$\,dBi gain. A detailed description of the measurement equipment in use can be found in~\cite{vehmas}.

\begin{figure}[!t]
  \centering
  \includegraphics[scale=0.48]{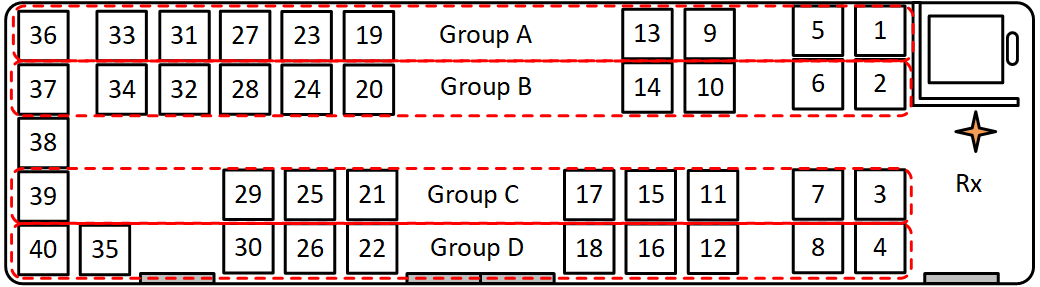}
  \caption{Schematic view of the bus with the seats numbered and divided into four groups. 
  }
  \label{fig:plan}
\end{figure}

\begin{figure}[!t]%
    \centering
    \subfloat[The upper transmitter position for the seat N\textsuperscript{\underline{o}}14.]
    {{\includegraphics[scale=0.9]{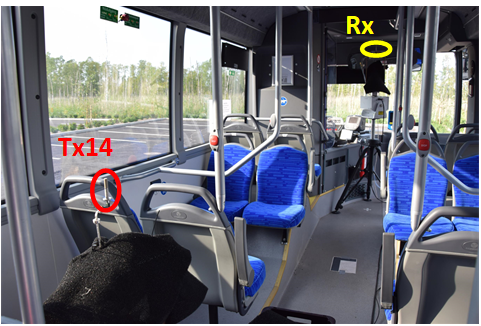} }}
    \quad
    \subfloat[The lower transmitter position for the seat N\textsuperscript{\underline{o}}24.]
    {{\includegraphics[scale=0.9]{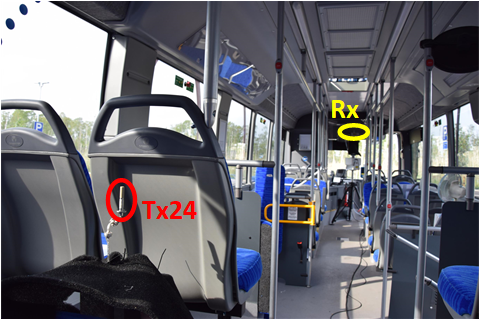} }}
    \caption{Photographs of the measurement setup.}%
    \label{fig:photos}
    \vspace{-0.5cm}
\end{figure}

We replicate a realistic consumer scenario through a specific device deployment inside the vehicle. As shown in Fig.~\ref{fig:plan}, the receiver is located in the front part of the bus at a height of $2$\,m (corresponds to a wireless access point). The transmitters are placed at (i) upper positions at the level of $1.2$\,m, which corresponds to, e.g., devices located near or attached to the owner's head (Fig. \ref{fig:photos}, a) and (ii) lower positions at a relative height of $0.7$\,m, e.g., the location of a mobile hand-held device (Fig. \ref{fig:photos}, b). 

The measurements have been performed for approximately one minute each to record ten samples of the channel impulse response (CIR). In total, we collected $72$ measurement data sets for the upper and lower positions of the transmitters. We exclude some of the lower transmitter locations (i.e., seats N\textsuperscript{\underline{o}}5-8 and N\textsuperscript{\underline{o}}27-30, see Fig.~\ref{fig:plan}) due to a convex shape of the bus floor around the wheels.

\begin{figure}[!t]
  \centering
  \includegraphics[scale=0.6]{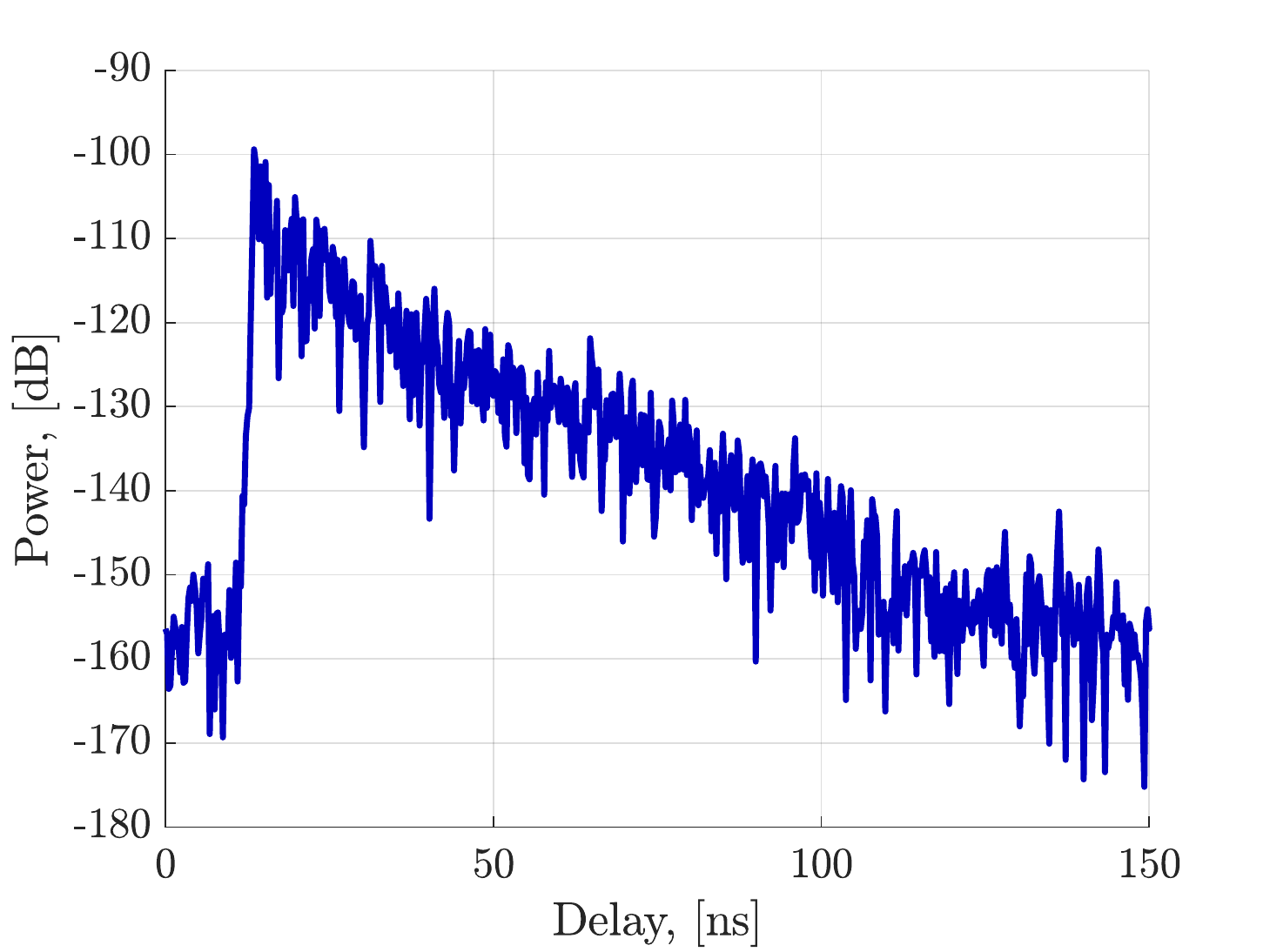}
  \caption{An example of the measured power delay profile for the transmitter position N\textsuperscript{\underline{o}}14 shown in Fig.~\ref{fig:photos}(a).}
  \label{fig:PDP}
  \vspace{-0.2cm}
\end{figure}

\vspace{7pt}
\section{Measurement Results and Analysis} 

In this section, we present the results of our measurement campaign and construct a logarithmic path loss model based on the received power data sets for different groups of seats. 

Fig.~\ref{fig:PDP} illustrates an example of the power delay profile (PDP) for the upper transmitter location N\textsuperscript{\underline{o}}14. The maximum value of the measured power is $-99.4$\,dB at $13.5$\,ns, which corresponds to the distance of $4.05$\,m between the transmitter and the receiver. 


Further, we derive the received power from the channel impulse response data as an integral value over all of the components and calculate the radiated power by compensating losses and gains introduced by the system elements. Using the difference between the radiated power and the received power and the antenna gains, we may obtain the sought path loss values. 


We subdivide our transmitters into four groups as illustrated in Fig.~\ref{fig:plan} and study the properties of these four groups separately. We construct a logarithmic model of the mean path loss as an increasing function of the distance between a transmitter and the receiver with a random fading term represented by a normally distributed random variable. Particularly, path loss values $L(d)$ can be approximated by the following expression:
\begin{equation}
 L(d) = \alpha + 10 \beta \log_{10}(d) + \chi(0,\sigma^2),
 \label{eq:logmodel}
\end{equation}
where $\alpha$ and $\beta$ are the propagation constant and the propagation exponent, respectively, and $\chi$ is a random Gaussian variable with zero mean and standard deviation $\sigma$. The parameters for all four groups are summarized in Table \ref{table:coefficients}.

\begin{figure}[!t]
	\centering
	\includegraphics[scale=0.64]{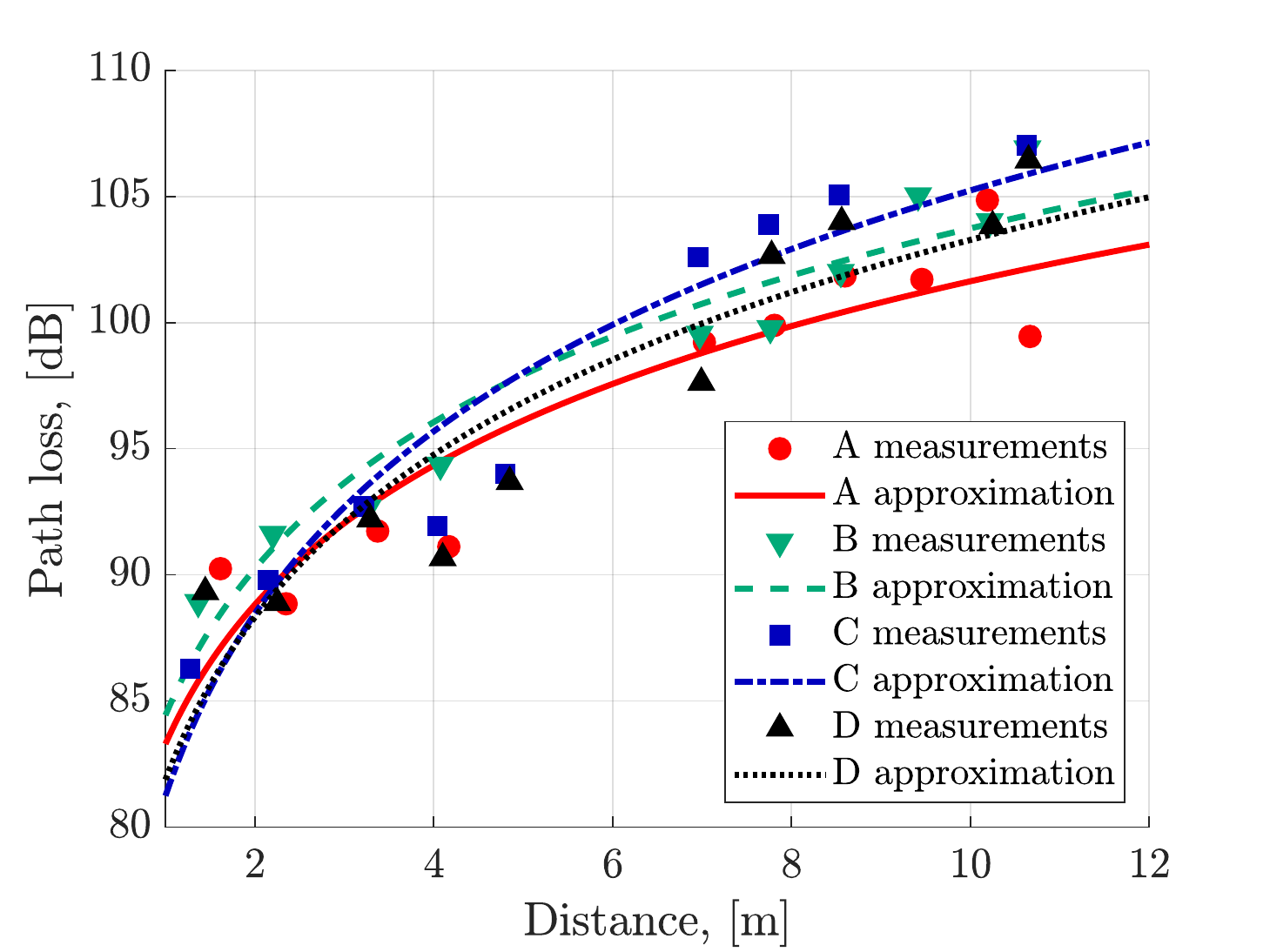}
	\caption{The path loss for the upper transmitter positions.}
	\label{fig:high_rows_PL}
\end{figure}

\begin{figure}[!t]
	\centering
	\includegraphics[scale=0.64]{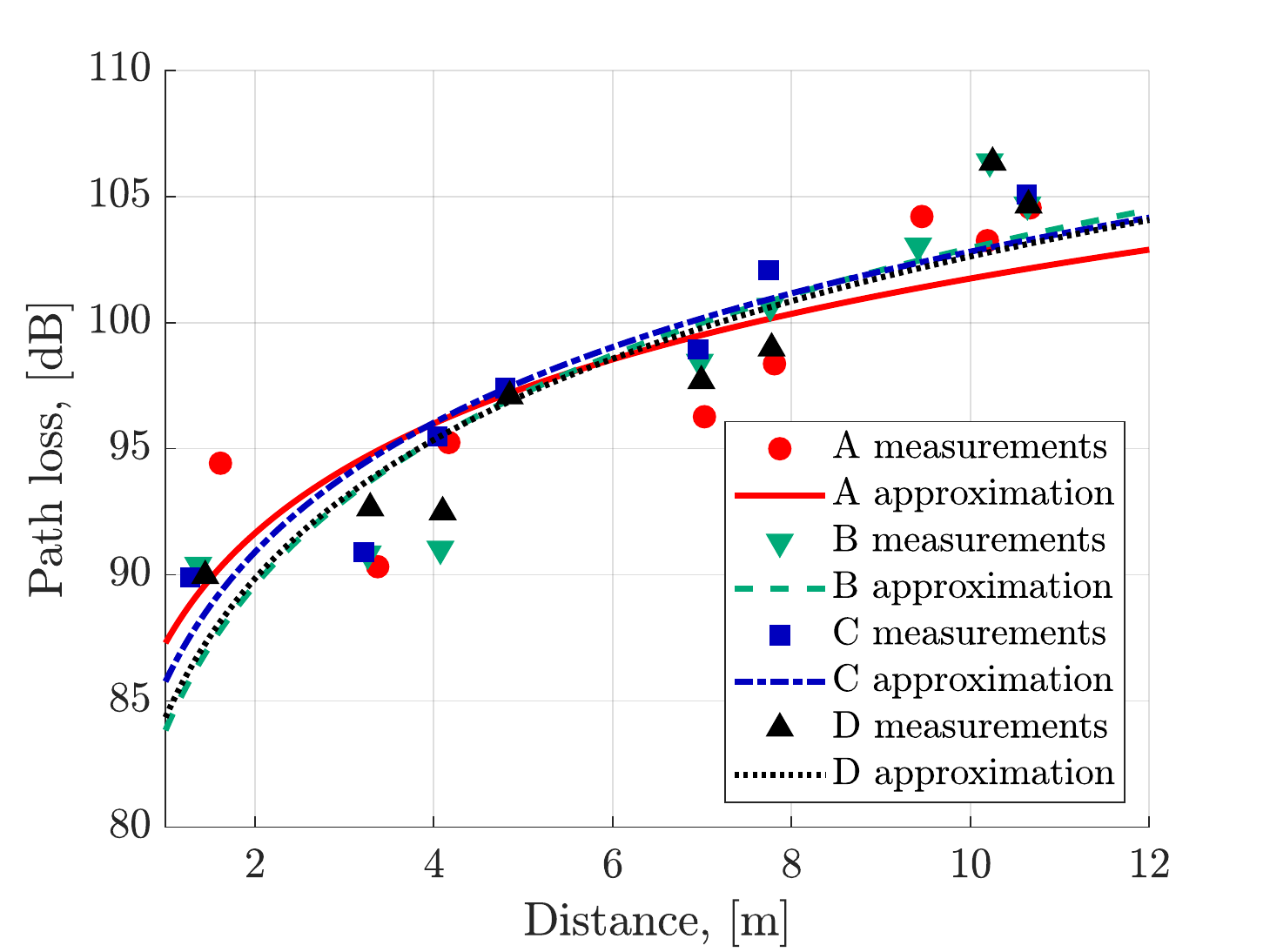}
	\caption{The path loss for the lower transmitter positions.}
	\label{fig:low_rows_PL}
\end{figure}

The path loss data for the upper and the lower transmitter positions are illustrated in Fig.~\ref{fig:high_rows_PL} and \ref{fig:low_rows_PL}, respectively. For the upper positions in Fig.~\ref{fig:high_rows_PL} and the distances shorter than $2.2$\,m, the path loss values of group ``A"  are slightly higher than those for the other regions, which is due to the wall behind the driver seat and less probable line-of-sight (LOS) connection. On the contrary, for the distances greater than $4$-$6$\,m, the path loss curve of group ``A" lies noticeably lower (up to $5$\,dB) than those of the other three groups, which results from the geometry of the bus and handrail locations (as can be seen in Fig.~\ref{fig:photos}a) that affect the probability of a LOS connection. The values of group ``D" are slightly different from those in ``A", which may be caused by the fact that the bus handrails are not symmetrical with respect to the receiver. 

In Fig.~\ref{fig:low_rows_PL}, we depict the estimated mean path loss values for the lower positions.  The measured values and the mean path loss curves for groups ``A", ``B", ``C", and ``D'' demonstrate similar behavior; a slight variation is caused by the seat locations in the bus. More detailed studies of these effects can be carried out with the use of ray-tracing simulations.

\begin{figure}[!t]
	\centering
	\includegraphics[scale=0.64]{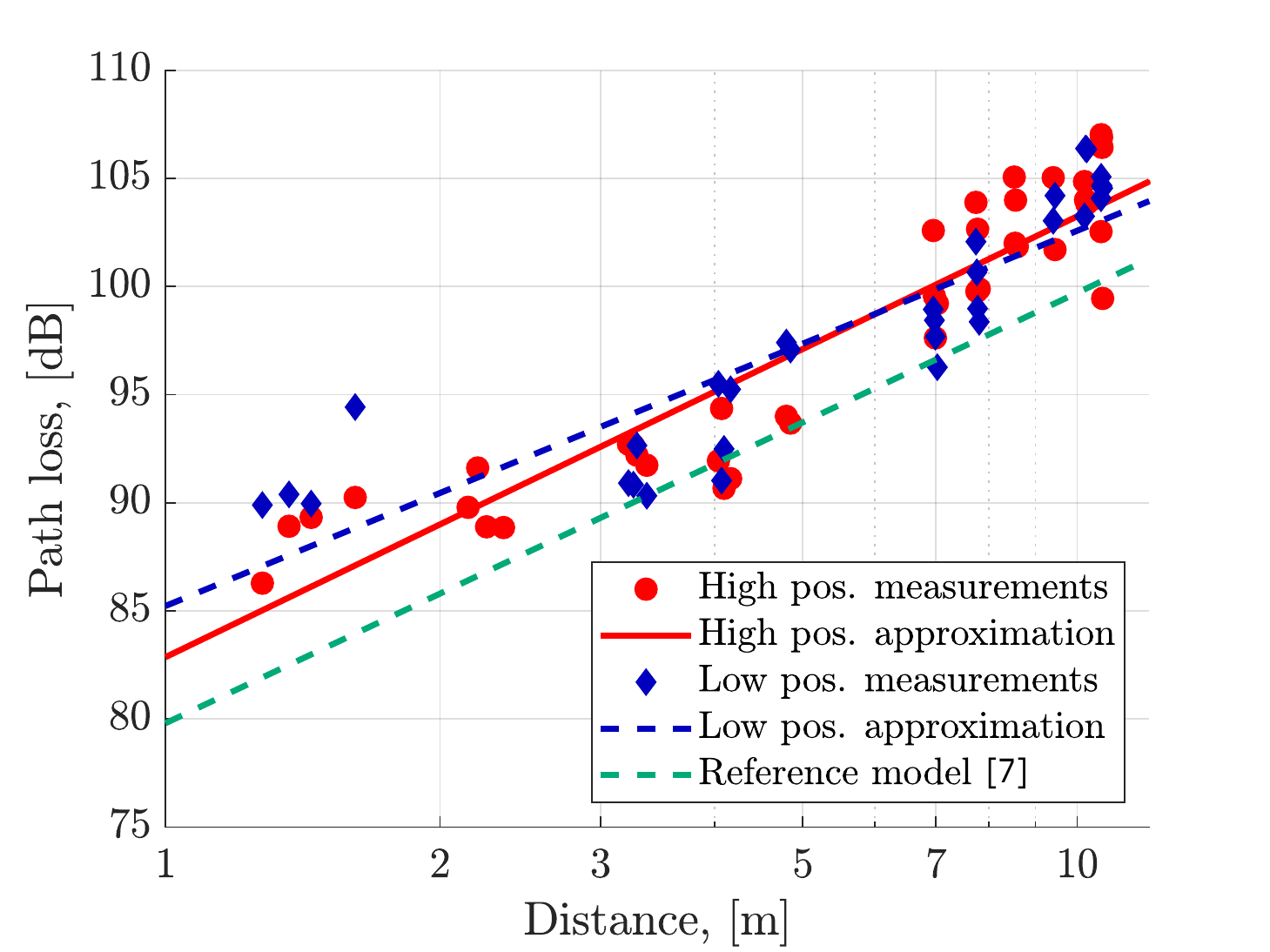}
	\caption{Comparison of the path loss logarithmic models for the upper and the lower positions (combined for all regions) with the results obtained in~\cite{chandra16_PL_60G_bus}.}
	\label{fig:comparison_PL}
\end{figure}

\begin{table}[!t]
	\centering
	\caption{Parameters of the proposed model.}
	\begin{tabular}{|c|c|c|c|c|}
		\hline
		\textbf{Region} & \textbf{Position} & \textbf{$\alpha$} & \textbf{$\beta$} & \textbf{$\sigma$} \\
		\hline
		\multirow{2}{*}{A} & Lower & 87.29 & 1.44 & 3.13  \\
		& Upper & 83.29 & 1.83 & 2.22 \\
		\hline
		\multirow{2}{*}{B} & Lower & 83.83 & 1.91 & 2.88  \\
		& Upper & 84.43 & 1.92 & 1.67 \\
		\hline
		\multirow{2}{*}{C} & Lower & 85.77 & 1.70 & 2.00  \\
		& Upper & 81.24 & 2.39 & 2.27 \\
		\hline
		\multirow{2}{*}{D} & Lower & 84.34 & 1.82 & 2.38  \\
		& Upper & 81.88 & 2.13 & 2.65 \\
		\hline
		\multirow{2}{*}{All} & Lower & 85.23 & 1.74 & 2.54  \\
		& Upper & 82.86 & 2.03 & 2.34 \\
		\hline
	\end{tabular}
	\label{table:coefficients}
\end{table}

\vspace{7pt}
\section{Discussion and conclusions}



In this paper, we summarize the results of a wideband measurement campaign carried out in an electric Linkker bus. We analyze two data sets: particularly, for (i) the upper and (ii) the lower antenna positions, which corresponds to a realistic deployment of devices in a dense public transportation environment. For the upper transmitter positions, our study shows that for longer distances all four groups of seats demonstrate different mean path loss values, which likely happens due to the blockage of the signal by the bus handrails. Moreover, the handrail locations are not symmetrical on both sides of the bus, which also causes the absence of symmetry for the corresponding groups. For the lower transmitter positions, the path loss curves behave similarly for the entire distance traveled. 

Combining the data for all four groups, we obtain the following expression for $L(d)$:
\begin{equation}
\left \{  \!\!
\begin{array}{l}
 L(d) = 85.2 + 17.4 \log_{10}(d) + \chi(0,6.5), \text{ lower positions},\\
  L(d) = 82.9 + 20.3  \log_{10}(d) + \chi(0,5.5), \text{ upper positions}.
 \end{array} 
 \right.
 \nonumber
\end{equation}
		
Analyzing the derived models, we also compare our results with the model reproduced from~\cite{chandra16_PL_60G_bus} (see Fig.~\ref{fig:comparison_PL}). We may observe that the path loss inside a typical city route bus is approximately $4$\,dB higher than that inside a touristic bus presented in~\cite{chandra16_PL_60G_bus}, which could stem from the different geometry and ceiling height. Despite non-identical antenna locations that also lead to small discrepancies, the results demonstrate a reasonable match. 

The models developed in the course of this study can become of a considerable benefit for subsequent calculations of the link budget and interference footprint studies. A more realistic scenario with the presence of human body blockage and analysis of the data sets for directional measurements are a part of our future work, together with the thorough ray-tracing simulation studies.

\vspace{7pt}
\section*{Acknowledgment}

The authors are particularly grateful to HSL (Helsinki Region Transport Authority) for providing the Linnker bus for the measurement campaign. Vasilii Semkin would like to thank Finnish Cultural Foundation (Suomen Kulttuurirahasto) for the support. This work is supported in part by the Academy of Finland, project WiFiUS. The work of Olga Galinina is supported by a personal Jorma Ollila grant from Nokia Foundation, by the Finnish Cultural Foundation, and by a Postdoctoral Researcher grant from the Academy of Finland.

\vspace{7pt}

\bibliographystyle{IEEEtran}
\bibliography{refs}

\end{document}